\documentclass[12pt,epsfig]{iopart}
\usepackage{graphicx}
\usepackage{epsfig}

\begin{document}

\title{$\phi$ meson production in d + Au collisions at
$\sqrt{s_{NN}}$ = 200 GeV}

\author{Dipali Pal \it{for the PHENIX collaboration\footnote[1]{See Appendix for the full collaboration list.}}  }

\address{  Department of Physics \& Astronomy, Vanderbilt University,
Nashville, TN 37235, USA
}

\begin{abstract}

We present the preliminary results on  $\phi$ meson production in the 
$\phi \rightarrow K^{+}K^{-}$ and $\phi \rightarrow e^{+}e^{-}$ decay channels measured at 
mid-rapidity in $\sqrt{s}$ = 200 GeV d + Au collisions at RHIC by the 
PHENIX experiment. The transverse mass spectra were obtained in 
both channels. The extracted $\phi$ yields are found to be consistent
with each other. The results are compared to the measurements in Au + Au 
collisions at the same center of mass energy.

\end{abstract}




\section{Introduction}

The $\phi$ meson is an important probe for studying the chiral symmetry
restoration and strangeness production in relativistic
heavy ion collisions. The decay of the $\phi$ meson 
into dilepton ($e^{+}e^{-}$  or $\mu^{+}\mu^{-}$) and $K^{+}K^{-}$
channels can probe the final state of the collision differently.
The dileptons have insignificant interactions with the medium, whereas
the kaons can scatter until freeze-out. The presence of a hot and
dense medium can modify the spectral properties (mass and/ or width) of 
the $\phi$ mesons {\cite{phi1}}.
However, there are predictions of $\phi$ width modifications in 
the cold nuclear matter as well {\cite{phi2}}. In general, it is quite 
important to investigate the evolution of the $\phi$ meson spectral
properties from cold  to hot 
nuclear matter created in d + Au and Au + Au collisions, respectively.

The $\phi$ meson consists of $s\bar{s}$ quarks. It is, therefore, sensitive
to the strangeness production in high energy collisions. The production 
mechanism of strangeness in heavy ion collisions can be investigated through
the measurement of the particle yields. The evolution of 
$\phi$ production with system sizes can be studied by measuring the yields 
in different centrality classes in Au + Au and d + Au collisions.

The PHENIX experiment at RHIC has measured $\phi$ mesons in both Au + Au and d + Au collisions. We shall mainly concentrate on the preliminary
results of $\phi$ meson measurement in d + Au collisions and its comparison
with Au + Au collisions at $\sqrt{s_{NN}}$ = 200 GeV.

\section{Experimental setup and data analysis}

The results described here are obtained using the two central
arm spectrometers {\cite{phenix}}, located East and West of the beam line at mid-rapidity with $\pi/2$ radian azimuthal coverage each. 

The collision vertex and centrality are measured by the  Beam Beam Counter. The momentum of each track is determined
by the drift chamber along with a layer of pad chamber. Tracking and
pattern recognition is accomplished by three layers of pad chamber. 

The
$\phi \rightarrow K^{+}K^{-}$ reconstruction 
uses kaons identified with high resolution Time-of-flight (TOF) wall at the
East arm within $|\eta|$ $<$ 0.35 and $\Delta\phi ~ \sim \pi/8$ and
the Lead Scintillator (PbSc) arrays located at both East and West spectrometer arms.
The TOF identifies kaons within $0.3 < p (GeV/c) < 2.0$ while the PbSc provides $\pi / K$ separation within a momentum range of $0.3 <  p (GeV/c) <  1.0$. 

The electrons, used for $\phi$ meson reconstruction
in $e^{+}e^{-}$ channel, are primarily identified by the Ring Imaging Cerenkov Detector. Further identification was provided by requiring the energy in the electromagnetic calorimeter to match the measured momentum of the tracks.

We have analyzed 62 $\times$ $10^{6}$ minimum-bias d + Au events within $|z_{vertex}| < $ 30 cm for $\phi \rightarrow K^{+}K^{-}$ analysis and 31 $\times$ $10^{6}$ single-electron triggered events for $\phi \rightarrow e^{+}e^{-}$ analysis.

\section{$\phi$ meson reconstruction}

The reconstruction of $\phi$ mesons takes place in two steps. First, we 
combine oppositely charged tracks (kaons for $\phi \rightarrow K^{+}K^{-}$
and electrons for $\phi \rightarrow e^{+}e^{-}$) to form unlike sign 
invariant mass spectrum which has some combinatorial background. In the second
step, we estimated the combinatorial background by event mixing technique.

\begin{figure}[]
\includegraphics[width=1.0\linewidth]{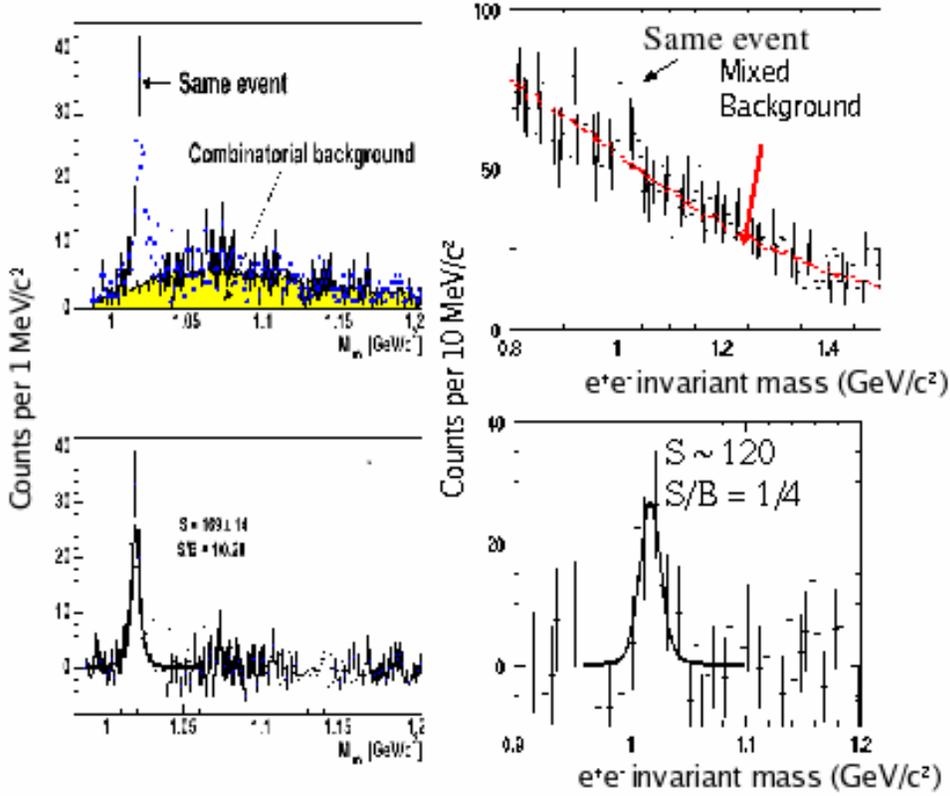}
\caption{\label{phi_inv} $\phi \rightarrow K^{+}K^{-}$ (left) $\phi \rightarrow e^{+}e^{-}$ (right) invariant mass spectra. In each plot, the upper panel shows
the same event and combinatorial invariant mass distribution while the lower panel exhibits the subtracted mass spectrum.
}
\end{figure}

For the $\phi \rightarrow K^{+}K^{-}$ analysis, we combined all $K^{+}$'s 
from one event with all $K^{-}$'s from the ten other events of the same 
centrality and vertex class. By limiting ourselves to a combinatorial 
background that does not exceed more than ten times 
the statistics of the actual distribition, the errors for fluctuations are ensured to be Poisson {\cite{soltz}}.  The validitr of this event mixing technique is 
confirmed with like sign distributions. Finally, the unlike sign 
 mixed event mass distribution is normalized to the 
measured 2$\sqrt{N_{++}N_{--}}$. The systematics associated with the 
normalization procedure is studied by normalizing the same event 
unlike sign distributions to the mixed events within different invariant
mass ranges above $\phi$ mass range. Typically, we use four different normalization ranges namely, $M_{K^{+}K^{-}}$ $<$ 1.1, 1.15, 1.2 and 1.25 GeV/$c^{2}$.
The systematic error in the extracted $\phi$ yield from normalization is
estimated to be around 4\%. The invariant mass spectrum of the $\phi$ mesons
in $K^{+}K^{-}$  decay channel is shown in Fig.~\ref{phi_inv}.

For the $\phi \rightarrow e^{+}e^{-}$ analysis extra care is taken to
form the mixed event background because the event sample used for 
di-electron analysis is triggered. The minimum-bias events are
used to form the mixed pair, where one of the electrons
is required to pass the trigger requirements. Additionally, as in $K^{+}K^{-}$
pairing, events used to form the mixed pairs are required to have similar centralities and vertex positions. The normalization of the mixed event
background is done by matching the data to the background in a 
sideband region between 850 and 950 MeV/$c^{2}$ below the
signal and 1100 - 1200 MeV/$c^{2}$ above the signal. Other methods of normalization are 
used to calculate the systematic error from this procedure. 

The invariant mass spectra are fitted with relativistic Breit Wigner distributions convolved with a Gaussian $\phi$ mass resolution function to study the
line shape of this vector meson.
The preliminary analysis shows the  $\phi$ meson spectral shape to be
consistent with the Particle Data Book.

\section{Corrections}

In order to make the transverse mass spectra, we corrected the $\phi$ meson 
yields for detector acceptance and reconstruction efficiency by Monte-Carlo simulation. This is carried
out by generating single $\phi$ mesons with exponential $p_{T}$ distribution
assuming an inverse slope of $T ~= ~ 320$ MeV and processing these simulated
events through PHENIX detector simulation chain that makes the simulated
data to be tuned with realistic detector responses. Finally, we calculate
the acceptance correction factors defined as
\begin{equation}
\epsilon = N_{\phi}^{Generated} / N_{\phi}^{Reconstructed}
\end{equation}
for different $m_{T}$ bins.

In addition, we apply the corrections associated with triggering efficiency and
efficiencies related with experimental run period to obtain the corrected
$\phi$ yields.

\section{Transverse mass spectra and yields}

\begin{figure}[t]
\includegraphics[width=1.0\linewidth]{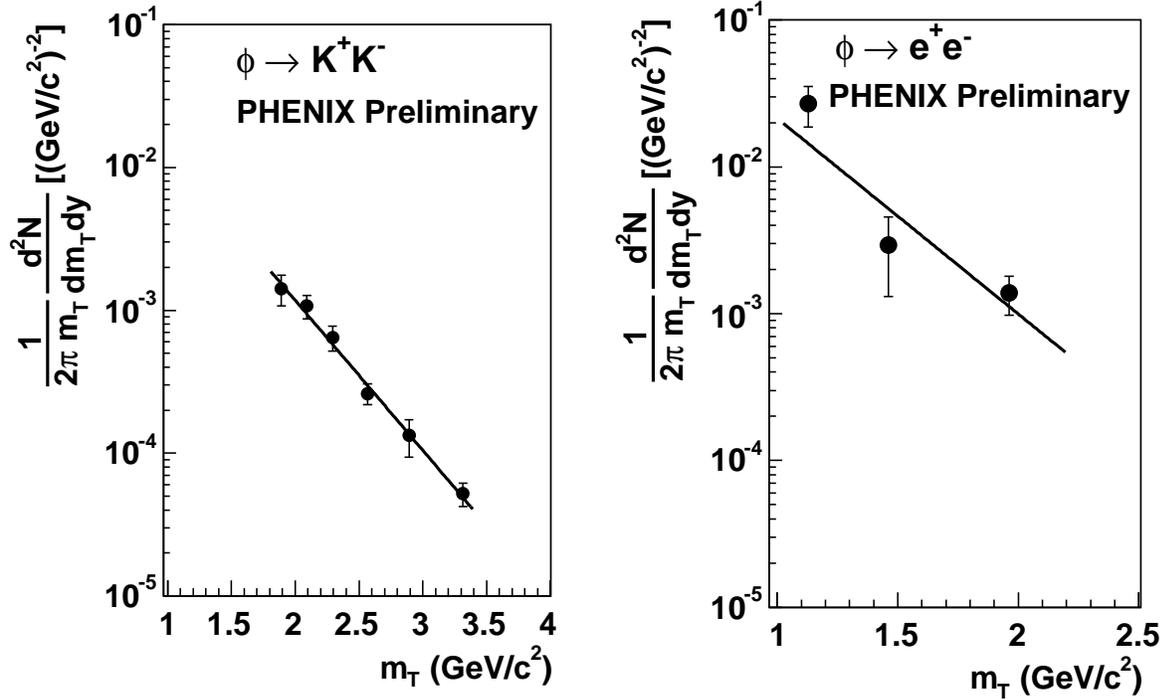}
\caption{\label{mtphi} Transverse mass distributions of $\phi$ mesons
reconstructed in $K^{+}K^{-}$ (left) and $e^{+}e^{-}$ decay channels.
}
\end{figure}

The minimum bias $m_{T}$ distributions of the $\phi$ mesons are reconstructed
by dividing the whole dataset into different $m_{T}$ bins. The minimum bias
transverse 
mass distributions of $\phi$ mesons 
are shown in Fig.~\ref{mtphi}. The transverse mass spectra in $K^{+}K^{-}$ 
decay channel is shown in the left hand side while $\phi \rightarrow e^{+}e^{-}$ $m_{T}$ spectrum is shown on the right. Each spectrum is fitted with exponential
function in $m_{T}$:
\begin{equation}
\frac{1}{2 \pi m_{T}} \frac{d^{2}N}{dm_{T}dy} = \frac{dN/dy}{2 \pi T (T + M_{\phi})} e^{-(m_{T} - m_{\phi})/T}
\end{equation}
where dN/dy and T are extracted from the fitting as two free parameters.

The yield parameters, dN/dy and T are shown in Table~\ref{dndyt} including
the statistical and systematic errors. The preliminary data shows the
consistency in the $\phi$ yield, dN/dy and inverse slope, T in both channels.
It may be useful here to mention that a factor of 2 - 4 discrepancy in the $\phi$ yield in kaon and $\mu^{+}\mu^{-}$ decay channels
was observed by NA49 and NA50 experiments {\cite{na4950}} respectively in
Pb + Pb collisions at $\sqrt{s_{NN}}$ = 17.27 GeV; the yield in di-muon channel was higher. The possible medium effects which might had caused this 
discrepancy is unlikely to be present in case of d + Au collisions which
produce cold nuclear matter. The consistency in the $\phi$ yield in d + Au
collisions is therefore quite expected. The $\phi$ meson yield in di-electron 
channel in Au + Au collisions at $\sqrt{s_{NN}}$ = 200 GeV was 
mesured by the PHENIX {\cite{dsmqm}} on the basis of 2001 Au + Au run at RHIC. But, the statistical and systematic errors in that preliminary measurement was too high to observe any
discripency compared to $K^{+}K^{-}$ decay channel. However, this issue
will be readdressed in 2003-2004 high luminosity Au + Au collisions where
the PHENIX experiment has recorded 1.5 $\times$ $10^{9}$ events.

\begin{table}
\caption{\label{dndyt} dN/dy and T of $\phi$ mesons measured in $K^{+}K^{-}$
and $e^{+}e^{-}$ decay channels in d + Au collisions at RHIC.}
\vskip 0.5cm
\begin{tabular}{|c|c|c|}\hline
Decay Channel & dN/dy & T\\
&&(MeV)\\\hline\hline
&&\\
$\phi \rightarrow K^{+}K^{-}$ & 0.047 $\pm$ 0.009 (stat) $\pm$ 0.009 (syst)
                              & 414 $\pm$ 31 (stat) $\pm$ 23 (syst)\\
$\phi \rightarrow e^{+}e^{-}$ & 0.056 $\pm$ 0.015 (stat) $\pm$ 0.028 (syst)
                              & 326 $\pm$ 94 (stat) $\pm$ 173 (syst)\\
&&\\\hline
\end{tabular}
\end{table}

\section{System-size dependence}

\begin{figure}[t]
\includegraphics[width=1.0\linewidth]{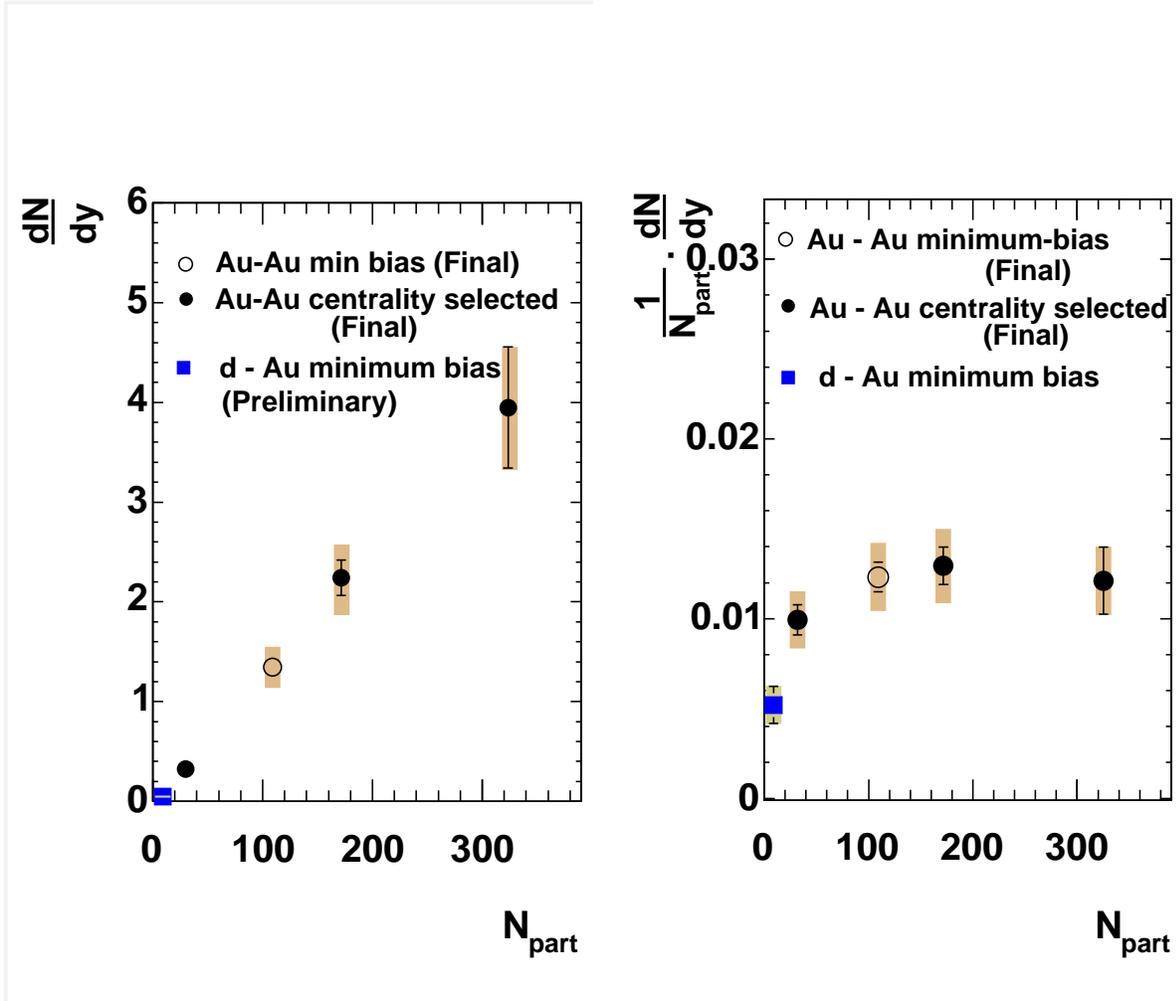}
\caption{\label{compare} $\phi$ yield (in $K^{+}K^{-}$ decay channel) in d + Au and Au + Au collisions
as a function of number of participants. The left hand side shows
dN/dy vs. $N_{part}$ while the right hand side presents dN/dy per participant
vs. $N_{part}$. The statistical error bars are shown by thin 
black lines whereas the systematic errors are indicated by thick brown lines.
}
\end{figure}

The $\phi$ meson yield in Au + Au collisions has also been measured
by the PHENIX experiment {\cite{phiprc,dsm}}. We compare the preliminary 
$\phi$ yield in
minimum bias d + Au collisions in $K^{+}K^{-}$ decay channel with the same in centrality selected
Au + Au collisions to measure the rate of $\phi$ production with
varying system sizes. The Fig.~\ref{compare} presents system size dependence 
of $\phi$ yield from d + Au to Au + Au collisions. Here we plot the 
variation of $\phi$ yields with number of participants measured by 
a Monte-Carlo calculation based on the Glauber model{\cite{glauber}}. The left hand side shows dN/dy as a function of number of participants indicating
an increase in dN/dy from d + Au minimum-bias to Au + Au collisions. The
right hand side plot, however, presents the variation of dN/dy per participant 
with number of participants. This shows a sharp increase in dN/dy per participant as we make a transition from d + Au minimum-bias to Au + Au peripheral (40 - 92\%) and then a saturation in case of Au + Au collisions.

\begin{figure}[t]
\includegraphics[width=1.0\linewidth]{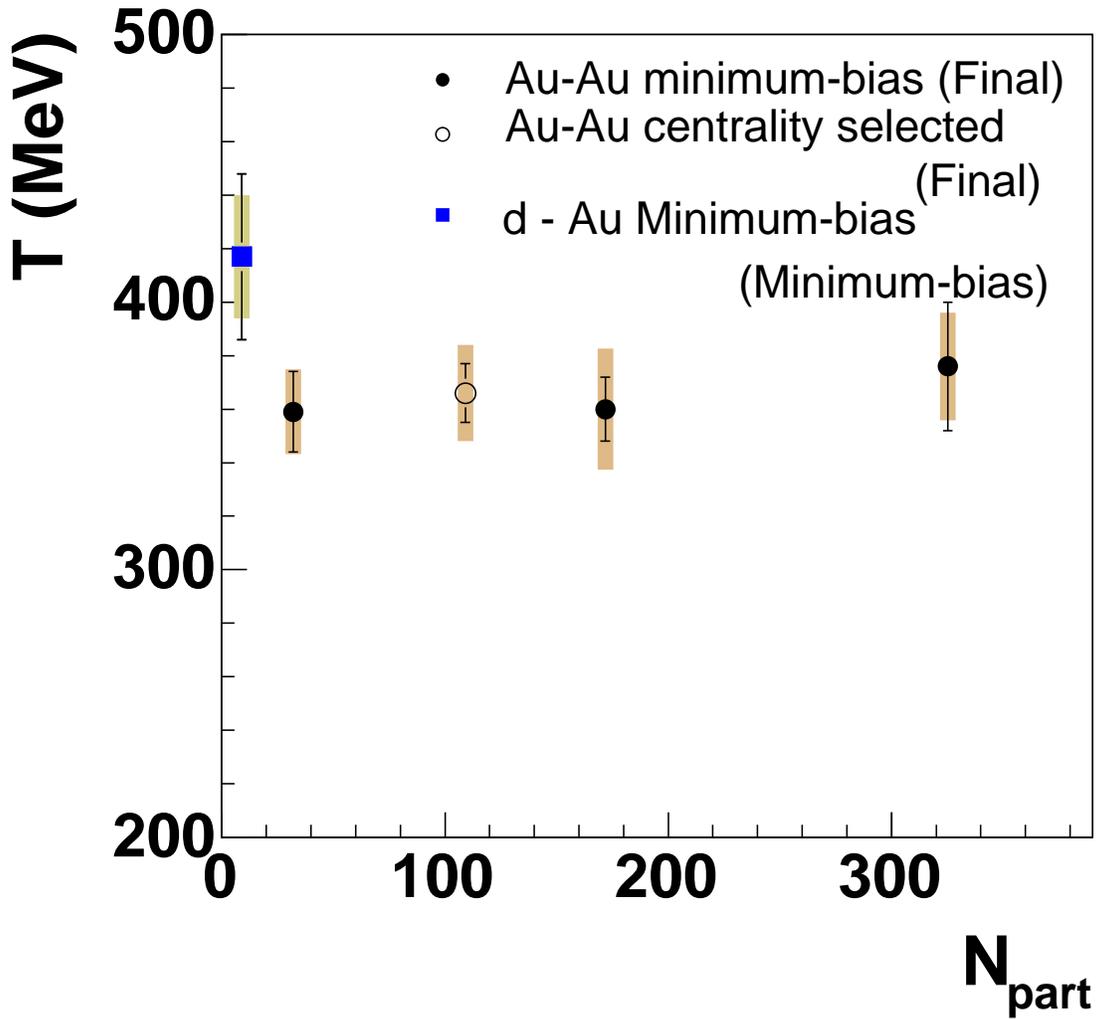}
\caption{\label{inverse} $\phi$ meson inverse slope (in $K^{+}K^{-}$ decay channel) in d + Au and Au + Au collisions
as a function of number of participants. The statistical error bars are shown by thin
black lines whereas the systematic errors are indicated by thick brown lines
}
\end{figure}

The variation of the inverse slope, T with $N_{part}$ is shown in Fig.~\ref{inverse}. We observe that the inverse slope of $\phi$ in $K^{+}K^{-}$ channel in Au + Au and d + Au collisions are consistent within statistical and systematic error bars. 

\section{Summary}

The PHENIX experiment at RHIC has measured $\phi$ meson yields both in
$K^{+}K^{-}$ and $e^{+}e^{-}$ decay channels in d + Au collisions at
$\sqrt{s_{NN}}$ = 200 GeV. The yield of the $\phi$ meson in both channels
are found to be consistent within errors. We have made a systematic comparison
between $\phi$ yield in d + Au and Au + Au collisions at $\sqrt{s_{NN}}$ = 200 GeV. We observe a sharp increase in dN/dy of the $\phi$ per participant from
d + Au to Au + Au collisions.


\section*{References}


\begin{thebibliography}{10}
\bibitem{phi1} K. Haglin, these proceedings.
\bibitem{phi2} F. Klingl, T. Waas and W. Weise, Phys. Lett. {\bf{B431}} 254 (1998).
\bibitem{phenix} K. Adcox {\it et al} (PHENIX Collaboration), Nucl. Instr. Meth. {\bf A499} 489 (2003).
\bibitem{soltz} R. A. Soltz, Ph.D. thesis, Department of Physics, Massachusetts Institute of Technology, 1994.
\bibitem{na4950} D. Rohrich, J. Phys. {\bf{G 27}}, 355 (2001).
\bibitem{dsmqm} Debsankar Mukhopadhyay for the PHENIX collaboration, Nucl. Phys. {\bf{A 715}}, 494 (2003).
\bibitem{phiprc} PHENIX Collaboration, S.~S.~Adler {\it et al.}, nucl-ex/0410012
\bibitem{dsm} D. Mukhopadhyay for the PHENIX collaboration, these proceedings.
\bibitem{glauber} S.~S.~Adler {\it et al.}, Phys. Rev. Lett. {\bf {91}} {072303} (2003).
\end{thebibliography}
\end{document}